%% file: main.tex
\documentclass[submission,copyright,creativecommons]{eptcs}
\providecommand{\event}{ICLP 2024} 

\usepackage{iftex}

\ifpdf
  \usepackage{underscore}         
  \usepackage[T1]{fontenc}        
\else
  \usepackage{breakurl}           
\fi

\usepackage[utf8]{inputenc}
\usepackage[english]{babel}
\usepackage{url}
\usepackage{amssymb}
\usepackage{amsmath}
\usepackage{graphicx}
\usepackage{amsthm}

\input{macro}

\input{HI-Maude_Listings}
\newtheorem{definition}{Definition}[section]

\newtheorem{property}{Property}[section]

\begin{document}

\title{Modular Stochastic Rewritable Petri Nets}

\author{Lorenzo Capra
\institute{
	{Dipartimento di Informatica}\\ {Universit{\`a} degli Studi di Milano}, Italy
}
}

\def\titlerunning{Modular Stochastic Rewritable Petri Nets}
\def\authorrunning{Lorenzo Capra}

\maketitle

\input{abstract.tex}

\section{Introduction}
\label{sec:intro}
\input{intro}

\section{(Stochastic) PT nets, \texttt{Maude}, and demonstrative example}
\label{sec:backgr}
\input{background}

\paragraph{Running example: fault-tolerant production line}
\label{sec:exe}
\input{example}

\section{Modular Rewritable Stochastic PN: Symmetries and Lumpability}
\label{sec:rewPT}
\input{modrewPT}
\section{Getting the Lumped CTMC generator from RwSPT}
\label{sec:CTMC}
\input{ctmc}

\section{Conclusion and Future Work}
We have developed a lumped Markov process for modular and rewritable Petri nets (RwPT), a flexible model of adaptive distributed systems.
RwPTs, which we construct and manipulate using a small set of algebraic operators, exhibit structural symmetries that result in an efficient quotient state-transition graph.
We have outlined a semi-automatic procedure for deriving the CTMC infinitesimal generator from the RwPT quotient graph. Future efforts will focus, on the one hand, on exploring orthogonal structured solutions and, on the other, on fully implementing the process and integrating it into graphical editors.
We aim to broaden the approach
to derive a lumped Markov process 
from any \texttt{Maude} specification.

\bibliographystyle{eptcs}
\bibliography{lc}

\end{document}

%% file: macro.tex
\newcommand{\Nat}{\ensuremath{\mathbb{N}}}

\newcommand{\Rel}{\ensuremath{\mathbb{R}}}

\newcommand{\bag}[1]{\ensuremath{Bag[#1]}}



%% file: HI-Maude_Listings.tex
\usepackage{listings}
\usepackage{xcolor}

\definecolor{delimiterColor}{HTML}{B65E47}
\definecolor{numberColor}{HTML}{FF0000}
\definecolor{commentColor}{HTML}{008000}
\definecolor{keyColor}{HTML}{002BFF}

\lstdefinelanguage{maude}
{
	breaklines=true,
	extendedchars=true,
	tabsize=2,
	columns=fullflexible,
	showtabs=false,
	showstringspaces=false,
	showspaces=false,
	showstringspaces=false,
	identifierstyle={\ttfamily},
	keywordstyle={\color{keyColor}},
	ndkeywordstyle={\color{keyColor}},
	stringstyle={\color{delimiterColor}},
	commentstyle={\color{commentColor}},
	ndkeywords={},
	keywords={pr, protecting, sort, sorts, op, ops, var, vars,eq, cq, ceq, crl, rl, mb, cmb, endfm, fmod, is, mod, endm, =, ==, =/=, ctor, ditto, Object, owise, Oid, prec, assoc, id, if, class, homod, endhom, eof, var, vars, eq, op, ops, pr, inc, protecting, including, ceq, is, tomod, endtom, sort, subsort, subsorts, to, endom, fmod, endfm, mod, endm, endtm, comm, gather, fth, endfth, format, metadata, memo},
	morecomment={[l]{***}},
	morecomment={[l]{---}},
}

%% file: abstract.tex
\begin{abstract}
Petri Nets (PN) are widely used for modeling concurrent and distributed systems, but face challenges in modeling adaptive systems. To address this, we have formalized 'rewritable' PT nets (RwPT) using \texttt{Maude}, a declarative language with sound rewriting logic semantics. Recently, we introduced a modular approach that utilizes algebraic operators to construct large RwPT models. This technique employs composite node labeling to outline symmetries in hierarchical organization, preserved through net rewrites.
Once stochastic parameters are added to the formalism, we present an 
automated process to derive a \emph{lumped} CTMC from the quotient graph generated by an RwPT.
\end{abstract}


%% file: intro.tex
Traditional formalisms such as Petri Nets, Automata, and Process Algebra do not make it easy for designers to define dynamic system changes. Several extensions inspired by the $\pi$-calculus and the Nets-within-Nets paradigm have been proposed, but they often lack suitable analysis techniques. Rewritable PT nets (RwPT) \cite{lcICDCIT22} is versatile formalism for analyzing adaptive distributed systems. RwPT is specified using the declarative language \texttt{Maude}, which adopts Rewriting Logic to offer both operational and mathematical semantics, creating a scalable model for self-adapting PT nets.
Unlike similar approaches (\cite{Barbosa11,RPN-Maude2016}),
the RwPT formalism provides data abstraction, is concise and efficient, and avoids the limitations posed by 'pushout' in Graph Transformation Systems. RwPT is an extension of GTS. Considering graph isomorphism (GI) when identifying equivalent states within the model dynamics is crucial to scaling up the model complexity.
Recent work has shown that GI is quasi-polynomial \cite{GIBabai}.
Graph Canonization (GC) involves finding a representative such that for any two graphs $G$ and $G'$, $G \simeq G' \Leftrightarrow can(G) = can(G')$. We developed a general canonization technique \cite{Capra:RP22} for use with RwPT, integrated into \texttt{Maude}. This technique works well for irregular models, but it is less effective for more realistic models.

In \cite{CAPRA-TCS2024}, we presented a technique for constructing comprehensive RwPT models using algebraic operators. The strategy is simple: leverage the modular characteristics of the models during analysis. By employing composite node labelling, we capture symmetries and sustain the hierarchical organization through net rewrites. A benchmark case study illustrates the effectiveness of our method.
In this paper, we demonstrate a procedure to derive a lumped Continuous-Time Markov Chain (CTMC) from the quotient graph formed by an RwPT model, after introducing stochastic parameters into the framework.

The potential of \texttt{Maude} as functional logic programming framework was discussed in \cite{Escobar2014} and recently in \url{https://logicprogramming.org/2023/02/extensions-of-logic-programming-in-maude/}. Although we do not consider free variables in terms, our work can be seen as an example of \emph{symbolic reachability} in which we use term rewriting through pattern matching (modulo normalization) instead of narrowing through unification. 
The use of narrowing to get a quotient state-transition system deserves further study.

%% file: background.tex

This section provides a concise overview of the (stochastic) PT formalism and emphasizes the key aspects of the 	\texttt{Maude} framework. For exhaustive information, we direct readers to the reference papers.

A \textit{multiset} (or \textit{bag}) $b$ on a set $D$ is a map $b: D \rightarrow \Nat$, where $b(d)$ is the \emph{multiplicity} of $d$ in $b$. We denote by $Bag[D]$ the set of multisets on $D$.
Standard relational and arithmetic operations can be applied to multisets on a component-by-component basis. 
%
A stochastic PT (or SPN) \emph{net}~\cite{ReisigPN,GSPN1993}
is a 6-tuple $(P,T,I,O,H,\lambda)$, where:
$P$, $T$ are finite, non-empty, disjoint sets holding the net's nodes (places and transitions, respectively);
$I,O,H:$ $T \ \rightarrow \bag{P}$ represent the transitions' \emph{input}, \emph{output}, and \emph{inhibitor} incidence matrices, respectively;
$\lambda : T \ \rightarrow \Rel^+$ assigns each transition a negative exponential firing rate. 
A PT net \emph{marking} (or state) is a multiset $m \in Bag[P]$.

The PT net dynamics
is defined by the \emph{firing rule}:
$t\in T$ is \emph{enabled} in $m$ if  
$I(t) \leq m$ and $\forall p \in P:  H(t)(p) = 0 \vee H(t)(p) > m(p)$. 
If $t$ is enabled in $m$ it may fire, leading to marking $m^\prime = m + O(t) - I(t)$.
We denote this: $m [ t \rangle m'$.
A PT-\emph{system} is a pair $(N,m_0)$, where $N$ is a PT net and $m_0$ is a marking of $N$.
The interleaving semantics of $(N,m_0)$ is specified by the \emph{reachability graph} (RG), an edge-labelled, directed graph $(V, E)$  whose nodes are markings. It is defined inductively: $m_0 \in V$; if $m \in V$ and
$m [ t\rangle m'$
then $m' \in V$, $m \xrightarrow{t} m' \in E$.
The timed semantics of a stochastic PT system is a CTMC isomorphic to the RG:
For any two $m_i, m_j \in V$, the transition rate from $m_i$ to $m_j$ is $r_{i,j} := \sum_{t : m_i [ t \rangle m_j} \lambda(t)$.


\vspace{-11pt}
\paragraph{Maude}
\texttt{Maude} \cite{maude07} is an expressive, purely declarative language characterized by a rewriting logic semantics \cite{rewlog03}. Statements consist of \emph{equations} and \emph{rules}. Each side of a rule or equation represents terms of a specific \emph{kind} that might include variables. Rules and equations have intuitive rewriting semantics, where instances of the left-hand side are substituted by corresponding instances of the right-hand side. The expressivity of \texttt{Maude} is realized through the use of matching modulo operator equational attributes, sub-typing, partiality, generic types, and reflection. A \emph{functional} module comprises only \emph{equations} and works as a functional program defining operations through equations, utilized as simplification rules. It details an \emph{equational theory} within membership equational logic \cite{membeqlog00}: Formally, a tuple $(\Sigma,E \cup A)$, with $\Sigma$ representing the signature, which includes the declaration of all sorts, subsorts, kinds\footnote{implicit equivalence classes defined by connected components of sorts (as per subsort partial order). Terms in a kind without a specific sort are \emph{error} terms.}, and operators; $E$ being the set of equations and membership axioms; and $A$ as the set of operator equational attributes (e.g., \texttt{assoc}). The model of $(\Sigma,E \cup A)$ is the \emph{initial algebra} $T_{\Sigma /E \cup A}$, which mathematically corresponds to the quotient of the ground-term algebra $T_{\Sigma}$. Provided that $E$ and $A$ satisfy nonrestrictive conditions, the final (or \emph{canonical}) values of ground terms form an algebra isomorphic to the initial algebra, that is, the mathematical and operational semantics coincide.

A \texttt{Maude} \emph{system module} includes \emph{rewrite rules} and, possibly, equations. Rules illustrate local transitions in a concurrent system. Specifically, a system module describes a generalized \emph{rewrite theory} \cite{rewlog03} $\mathcal{R}= (\Sigma,E \cup A,\phi,R)$, where $(\Sigma,E \cup A)$ constitutes a membership equation theory; $\phi$ identifies the frozen arguments for each operator in $\Sigma$; and $R$ contains a set of rewrite rules. 
A rewrite theory models a concurrent system: $(\Sigma,E \cup A)$ establishes the algebraic structure of the states, while $R$ and $\phi$ define the concurrent transitions of the system. The initial model of $\mathcal{R}$ assigns to each kind $k$ a labeled transition system (TS) where the states are the elements of $T_{\Sigma /E \cup A,k}$, and transitions occur as $ [t] \overset{[\alpha] }{\rightarrow} [t']$, with $[\alpha]$ representing \emph{equivalent} rewrites. The property of \emph{coherence} guarantees that a strategy that reduces terms to their canonical forms before applying the rules is sound and complete. A \texttt{Maude} system module is an executable specification of distributed systems. Given finite reachability, it enables the verification of invariant properties and the discovery of counterexamples.

%% file: example.tex
\begin{figure}
   \begin{center}    
    \begin{tabular}{rl}
        \fbox{\includegraphics[scale=0.33]{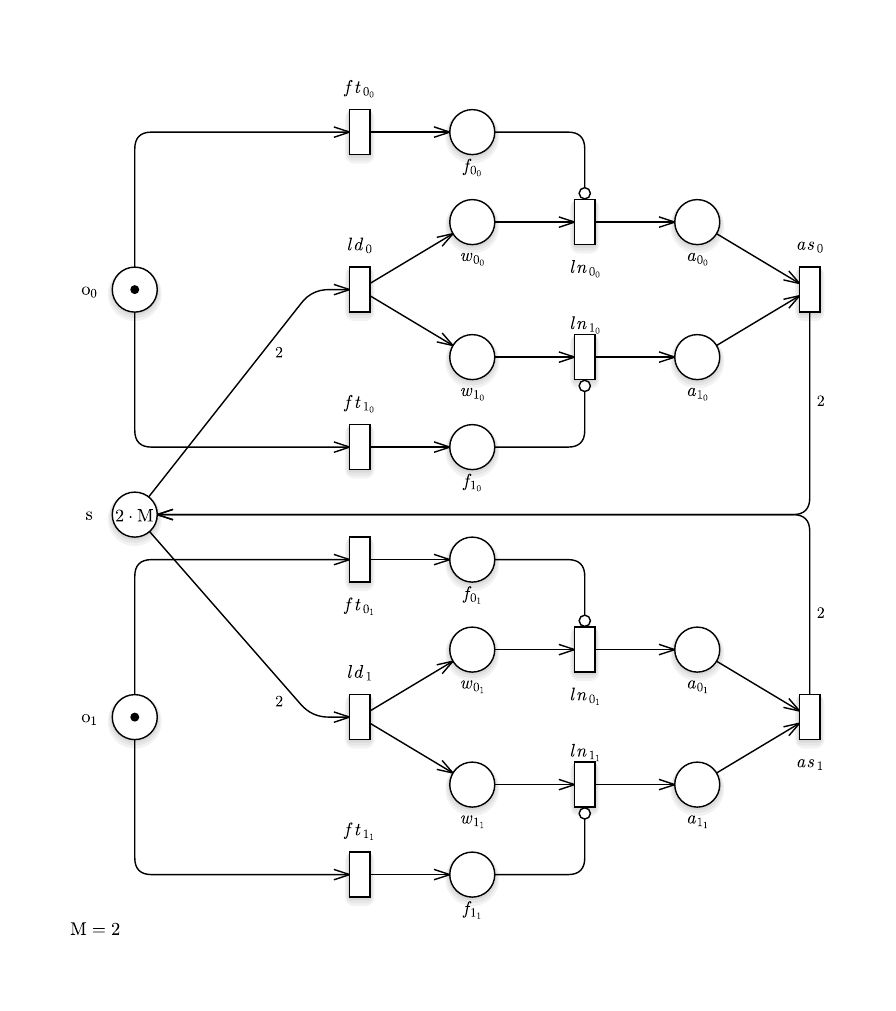}}&
        $\Rightarrow$
        \fbox{\includegraphics[scale=0.33]{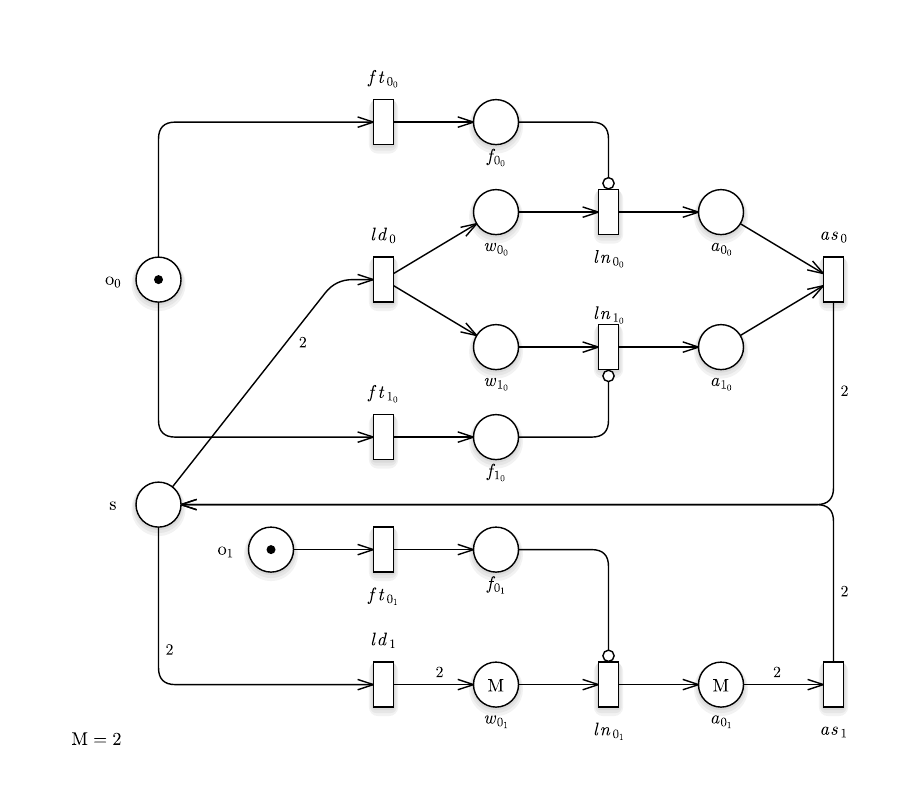}}
        $\Rightarrow$\\
        $~$&$~$\\
        $\Rightarrow$
        \fbox{\includegraphics[scale=0.33]{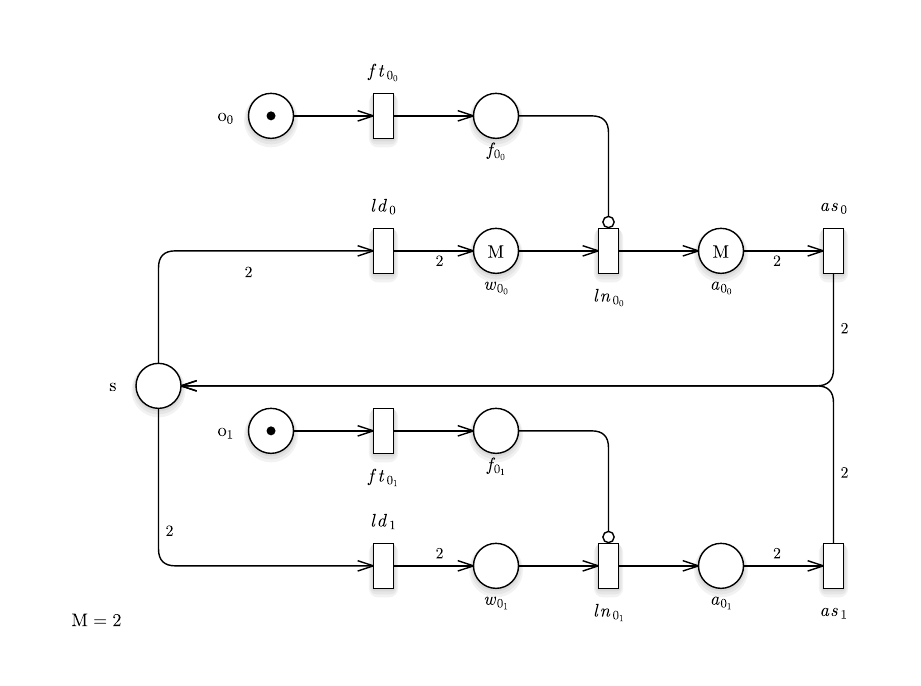}}&
        $\Rightarrow$
        \fbox{\includegraphics[scale=0.33]{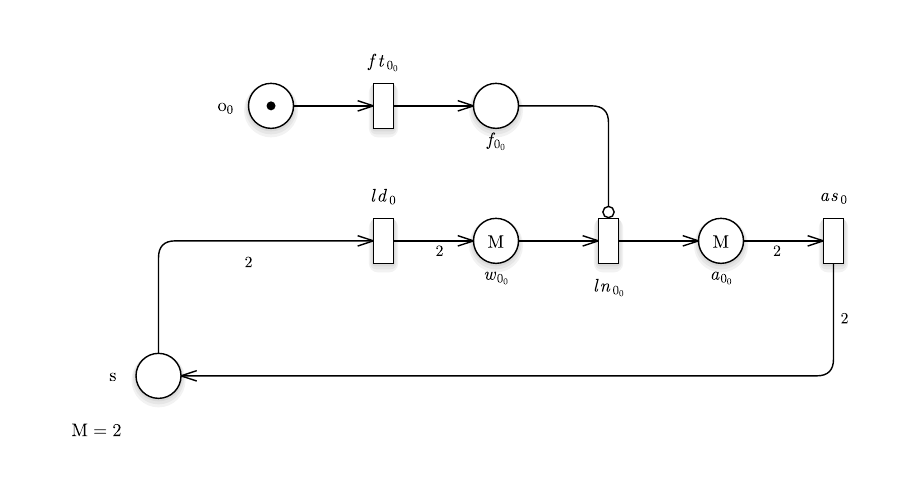}}
    \end{tabular}
    \end{center}
    \caption{One of the possible paths of the Gracefully Degrading Production System.}
    \label{fig:symPL-degarde2}
\end{figure}

As an illustrative example, we refer to the model of a distributed production system that gracefully degrades presented in \cite{CAPRA-TCS2024}. The system is composed of $N$ production lines (PL), each branching into $K$ fully interchangeable robots, that handle K raw items in parallel 
and assemble them into an artifact. In this study, $K = 2$. The items are loaded from a warehouse in an PL, $K$ at a time. 
A robot in a PL might fail, upon which
the PL restructures to continue functioning, but with reduced capacity.
The reconfiguration process involves moving items from the faulted branch of a PL to the remaining branch(es) to maintain the production cycle. Traditional PNs
are unable to model this operation. 
When a second fault occurs in a degraded PL, the system disconnects the PL. The leftover items are then relocated to the warehouse.
Figure \ref{fig:symPL-degarde2} shows the evolution of a system starting with two PLs. This scenario can be extended to $N$ PLs, each operating $K$ parallel robots,
that handle $K \cdot M$ items ($M$ being a third parameter of the model), 
denoted by the term \verb|NPLsys(N, K, M)|. 

%% file: modrewPT.tex
Rewritable stochastic PT nets (RwSPT) build upon the concept of \emph{modular} rewritable PT nets \cite{CAPRA-TCS2024} by linking negative exponential rates to the firing of PT transitions and the process of net rewrites.

The definition of RwSPT includes a hierarchy of \textbf{Maude} modules (e.g., \texttt{BAG}, \texttt{PT-NET}, \texttt{PT-SYSTEM}) described in  \cite{CAPRA-TCS2024}.
The \texttt{Maude} sources can be found in \url{https://github.com/lgcapra/rewpt/tree/main/modSPT}.
RwSPT uses structured annotations to underline the symmetry of the model. It features a concise place-based encoding to aid in state canonization and is based on the functional module \verb|BAG{X}|, which introduces multisets as a complex data type. The commutative and associative \verb|_+_| operator provides an intuitive way to describe a multiset as a weighted sum, for instance, \verb|3 . a + 1 . b|. The sort \verb|Pbag| contains multisets of places.
Each place label (a term of sort \verb|Plab|) is a nonempty list of pairs built of \verb|String| and a \verb|Nat|. Places are uniquely identified by their labels. These pairs represent a symmetric component within a nested hierarchy. Compositional operators annotate places incrementally from right to left: The label suffix represents the root of a hierarchy. 
For example, the 'assembly' place of line 1 in Production Line 2 would be encoded as:
\verb|p(< "a"; 0 > < "L"; 1 >)|.     
We implicitly describe net transitions (terms \verb |Tran|) through their incidence matrix (a 3-tuple of terms \verb|Pbag|) and associated tags. A tag includes a descriptive \verb|String|
and a \verb |Float| interpreted as a firing rate. 
The syntax is:
\verb|[I,O,H] -> << S, R >>|. 

Using the associative composition operator \verb"_;_" and the subsort relation \verb"Tran < Net", we can easily construct PT nets in a modular way. For example, we can depict the subnet containing the load transition ($ld$) and a robot ($ln_0$ ) as the \verb"Net" term in the listing below.

{\small\begin{lstlisting}[frame=single,language=maude]
  [2 . p(< "s" ; 0 >), 1 . p(< "w" ; 0 >) + 1 . p(< "w" ; 1 >), nilP] |->  << "ld", 0.5 >> ;
  [1 . p(< "w" ; 0 >), 1 . p(< "a" ; 0 >), 1 . p(< "f" ; 0 > ] |-> << "ln", 0.1 >>
\end{lstlisting}}

A \verb|System| term is the empty juxtaposition (\verb|__|) of a \verb|Net| and a \verb|Pbag| representing the marking. 
The conditional rewrite rule \verb|firing| specifies the PT firing rule
(notice the use of a matching equation :=).

{\small\begin{lstlisting}[frame=single,language=maude]
 vars N N' : Net . vars T  : Tran . var M : Pbag .
 crl [firing] : N M => N firing(T, M) if T ; N' := N  /\ enabled(T, N M) . 
\end{lstlisting}}


An RwSPT is defined by a system module that contains two constant operators, used as aliases:
\verb|op net : -> Net | and \verb|op m0 : -> Pbag |.
Two equations define their bindings.
This module includes a set $R$ of \verb|System| rewrite rules incorporating \verb|firing|.
We adopt interleaving semantics: Rewrites 
have an exponential rate (specified in the rule label but for \verb|firing|), so that for the state transition system it holds ($\subseteq$ is the subgraph relation):
$TS($\verb|net m0|, $\{$\verb|firing|$\}) \subseteq TS($\verb|net m0|, $R)$.

\vspace{-11pt}
\paragraph{Modularity, symmetries, and lumpability} We have provided net-algebra and net-rewriting operators \cite{CAPRA-TCS2024} with a twofold intention: to ease the modeler's task and to enable the construction and modification of large-scale models with nested components by implicitly highlighting their symmetry.
A compact \emph{quotient} TS is built using simple manipulation of node labels.
This approach outperforms that integrated into \texttt{Maude} \cite{Capra:RP22} and based on traditional graph canonization.

In this context, the identification of behavioral equivalences is reduced to a graph \emph{morphism}. PT system morphism must maintain the edges and the marking: In our encoding,
a \emph{morphism} between PT systems \verb|(N m)| and \verb|(N' m')| is a bijection $\phi \ :$ \verb|places(N)| $\rightarrow$ \verb|places(N')| such that, considering the homomorphic extension of $\phi$ on multisets, $\phi($\verb|N|$) = \ $\verb|N'| and $\phi($\verb|m|$) = \ $\verb|m'|.
Moreover, $\phi$ must retain the textual annotations of the place labels and the transition tags. If \verb|N'| = \verb|N| we speak of \emph{automorphism}, in which case $\phi$ is a permutation in the set of places.
We refer to a \emph{normal} form that principally involves identifying sets of automorphic (permutable) places:
Two markings \verb|m|, \verb|m'| of a net \verb|N| are said automorphic if there is an automorphism $\phi$ in \verb|N| that maps \verb|m| into \verb|m'|.
We denote this \verb|m| $\cong$ \verb|m'|. The equivalence relation $\cong$ is a congruence, that is, it preserves the transition firings and \emph{rates}.
\begin{definition}[Symmetric Labeling]
\label{def:modsym}
A \verb|Net| term is symmetrically labeled if any two maximal sets of places whose labels have the same suffix (possibly empty), which is preceded by pairs with the same tag, are permutable. A \verb|System| term is  symmetrically labeled if its \verb|Net| subterm is.
\end{definition}

\noindent In other words, if a \verb|Net| term \verb|N| meets definition \ref{def:modsym}, then for any two maximal subsets of places matching:

$P := \{$\verb|p(L' < w ; i > L)|$\}, \quad P' := \{$\verb|p(L'' < w ; j > L)|$\}$,

where: \verb|L, L', L'' : Plab, w: String, i, j : Nat| 

\noindent there is an automorphism $\phi$ such that $\phi(P) = P'$, $\phi(P') = P$,
which is extended as an identity to the rest\footnote{According to the definition of PT morphism, the prefixes \texttt{L'} and \texttt{L''} are consistent in the textual component.}.

If a \verb|System| term adheres to the previous definition, it can be transformed into a 'normal' form by merely swapping indices on the place labels (e.g., i $\leftrightarrow$ j), while still complying with definition \ref{def:modsym}. This normal form is the most minimal according to a lexicographic order within the automorphism class ($\cong$) implicitly defined by \ref{def:modsym}. In contrast to general graph canonization, there is no need for any pruning strategy or backtracking: A monotone procedure is used where the sequence of index swaps does not matter (see \cite{CAPRA-TCS2024}). Efficiency is achieved as the normalized form of the subterm \verb | Net | is derived through basic ``name abstraction``, where at each hierarchical level the indices of structured place labels continuously span from $0$ to $k \in \Nat$.  

The strategy involves providing a concise set of operators that preserve nets' symmetric labelling. This set includes \emph{compositional} operators 
and operators for \emph{manipulating} nets.
Rewrite rules require these operators to manipulate \verb|System| terms defined in a modular manner. 
Furthermore, rules must adhere to parametricity conditions that limit the use of ground terms \cite{CAPRA-TCS2024}.
Under these assumptions, we get a \emph{quotient} TS from a \verb|System|
term that
retains reachability and
meets strong bisimulation. 

Let $t,t',u,u'$ be (final) terms of sort \verb|System|, 
$r$ a \verb|System| rule $r : \ s \Rightarrow s'$. The notation $t \overset{r(\sigma)}{\Rightarrow} t'$
means there is a ground substitution $\sigma$ of $r$'s variables such that $\sigma(s) = t$ and $\sigma(s')= t'$.
\begin{property}
\label{prop:transition-corr}
Let
$t$ meet Definition \ref{def:modsym}.
If $t \overset{r(\sigma)}{\Rightarrow} t'$, then $\forall u, \phi, t \cong_{\phi} u$: $ u \overset{r(\phi(\sigma))}{\Rightarrow} u'$, $ t' \cong u'$ 
\end{property}

The TS quotient generated by a normal form $\hat{t}$ is obtained by applying the operator \verb|normalize| to the terms on the right side of the rewrite rules. When a \verb|System| undergoes a rewrite due to the \verb|firing| rule, the process only involves the marking subterm.
According to property \ref{prop:transition-corr} (firing preservation), because the morphism $\phi$ preserves the transition rates and the rules are parameterized, it is feasible to map the TS quotient of $\hat{t}$ onto an isomorphic "lumped" CTMC: In a Markov process's state space, an equivalence relation is considered "strong lumpability" if the cumulative transition rates between any two states within a class to any other class remain consistent. Despite the possibility of establishing a more stringent condition, namely "exact lumpabability," we focus on the aggregated probability.

%% file: ctmc.tex
A rewritable PT \texttt{system}  generates a transition system (TS) isomorphic to a lumped CTMC. 
However, the TS produced via the \texttt{show search graph} command of \texttt{Maude} embodies a \emph{parametric} CTMC: in line with the rewriting logic semantics, state transitions denote \emph{classes} of equivalent rewrites, meaning, PT firings that result in identical normalized markings or net rewrites that lead to isomorphic PT systems.

To obtain the CTMC generator, it is necessary to quantify instances that align with a specific state transition. Regrettably, the \texttt{Maude} system lacks a mechanism to determine the matches of a rewrite rule in the TS construction process.

  Our solution consists of first (automatically) generating a Transition System with states having a composite structure, which provides a detailed view of the equivalent rewrites that result in state transitions. Subsequently, using elementary parse to compute the cumulative rates of the lumped CTMC.
 Despite a redundant state representation, this method incurs an acceptable time overhead because it only involves \emph{normalized} states. 


When considering the term \verb|NPLsys(2,2,2)|, which aligns with the PT net at the top left of Figure \ref{fig:symPL-degarde2}, the resulting quotient TS comprises 295 states compared to the 779 states in the standard TS. State transitions often correspond to multiple matches: For instance, the initial state (the term above) includes two 'load' instances and four 'fault' instances that lead to markings with identical normal forms. Consequently, the combined rates are $2\cdot0.5$ and $4\cdot0.001$. 

\vspace{-10pt}
\paragraph{Experimental Evidence}
We showcase experimental validation of the method and a demonstration of standard performance indicators. Table \ref{tab:perf} displays the results of the \texttt{search} command to locate the final states.
We used Linux WSL on an 11th Gen. Intel Core i5 with 40GB RAM. The state spaces match those of the lumped CTMC. The analysis of large models is feasible solely by exploiting the model's symmetry.
Notice that the number of absorbing states in the TS quotient does not vary with $N$. 

\begin{figure}[htbp]
   \begin{center}    
    \includegraphics[clip,trim={2.1cm 2.7cm 2.2cm 2.2cm},width=0.7\columnwidth]{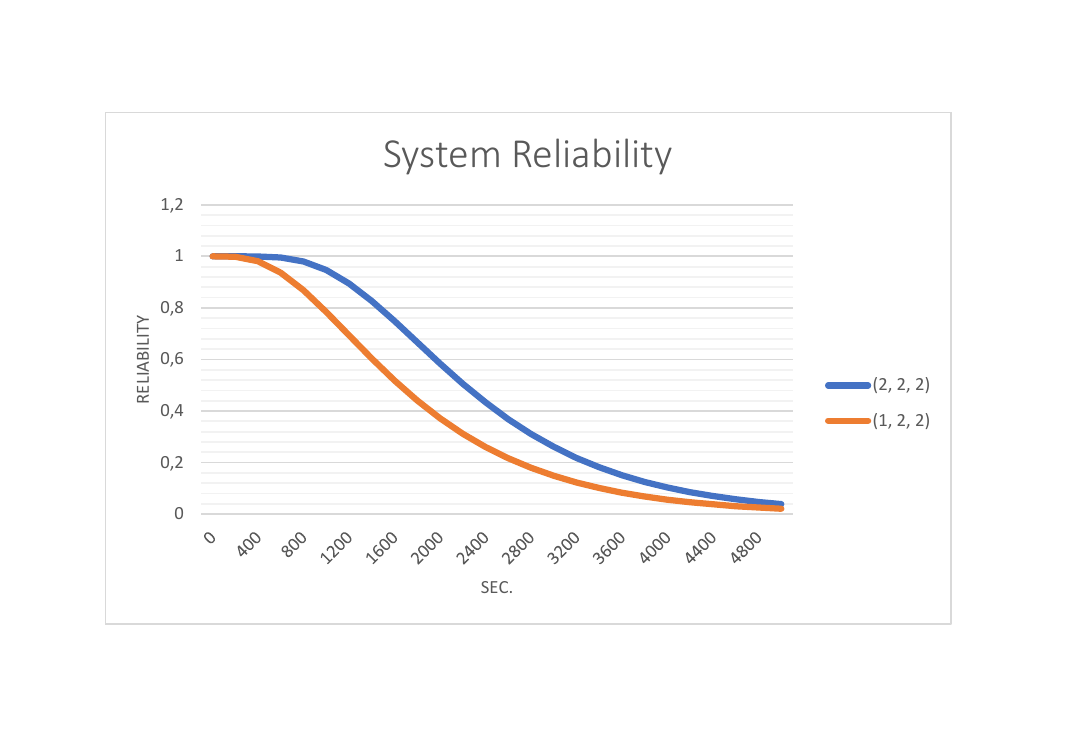}
    \end{center}
    \caption{System Reliability.}
    \label{fig:PlotRel}
\end{figure}

Figure
\ref{fig:PlotRel} shows the system \emph{reliability} (the complement of Time to System Failure distribution).
As expected, 
is decreases with time; additionally, the scenario that involves more replicas demonstrates
enhanced reliability.
The inflexion point at around time 800
represents the system's reconfiguration time. The increased execution time of the job (not reported) is a result of a system failure.
The overall trend is also noticeable when we look at larger values of \verb|N|. As \verb|N| increases, both reliability and throughput curves show significant improvements.
However, we observe an asymptotic trend when \verb|N| is greater than 6. Our interpretation is that beyond a certain point, the benefit of using a higher number of replicas is outweighed by the higher fault rate and the increased configuration overhead.

\vspace{-0.5cm}
\begin{table}[htbp]
\centering\small
\caption{Ordinary vs Quotient TS of \texttt{NPLsys(N,2,2)} \hspace{2pt} ${}^\dag$ \texttt{search} timed out after 10 h} 
\label{tab:perf}
\begin{tabular}{ |p{0.5cm}||p{2.5cm}p{1.5cm}||p{2.5cm}p{1.5cm}| }
\hline
 & \multicolumn{2}{c||}{Ordinary} & \multicolumn{2}{|c|}{Quotient} \\
\verb|N|   & states(final)  & time (sec) &  states(final) & time (sec) \\
\hline
\hline
1   &    60(2) & 0 &      42(2) & 0\\
\hline
2   &    779(4) & 0.1 &     295(2) & 0.1\\
\hline
3  &   6101(6) & 4.8  &  1059(2) & 0.9\\
\hline
4 &    37934(8) & 69 & 2764(2) & 3.6\\
\hline
5 &   204362(10) & 818 & 5970(2) & 10 \\
\hline
6 &   1000187(12) & 13930 & 11367(2) & 27\\
\hline
7 &  - & ${}^\dag$ & 19775(2) & 65 \\
\hline
8 &  - & ${}^\dag$ &  32144(2) & 186 \\
\hline
9 & - & ${}^\dag$ &  49554(2) & 569 \\
\hline
10 & - & ${}^\dag$ & 73215(2) & 2450 \\ 
\hline
\end{tabular}
\end{table}

